\documentclass[12pt,onecolumn]{article}
\usepackage[utf8]{inputenc}
\usepackage[T1]{fontenc}
\usepackage[english]{babel}

\usepackage{hyperref}

\usepackage{fancyhdr}

\usepackage{graphicx}
\usepackage{amsmath,amssymb}
\usepackage{cite}
\usepackage{authblk}

\title{
{\fontsize{14}{16}\selectfont\bfseries
	EQUILIBRIUM ADSORPTION OF HARD DISKS ON PATTERNED ADHESIVE
	SURFACES: A MONTE CARLO SIMULATION STUDY}	
}

\author[1]{N.~Kukarkin}
\author[1,2,*]{T.~Patsahan}

\affil[1]{\normalsize Yukhnovskii Institute for Condensed Matter Physics of the National Academy of Sciences of Ukraine, 1 Svientsitskii str., 79011, Lviv, Ukraine}
\affil[2]{\normalsize Institute of Applied Mathematics and Fundamental Sciences,  Lviv Polytechnic National University, 12 S.~Bandera Str., 79013 Lviv, Ukraine}
\affil[*]{\normalsize Corresponding author: tarpa@icmp.lviv.ua}

\affil[ ]{\small ORCID: 
	N.~Kukarkin, \href{https://orcid.org/0009-0005-8787-5639}{0009-0005-8787-5639}; 
	T.~Patsahan, \href{https://orcid.org/0000-0002-7870-2219}{0000-0002-7870-2219}}

\date{}

\begin{document}
	
	\maketitle
	
\begin{abstract}
Equilibrium adsorption of disk-like particles on patterned adhesive surfaces is studied using Monte Carlo simulations. The surface is represented as a two-dimensional plane with circular adhesive domains arranged either regularly or randomly, while the particles are modelled as hard disks. The interaction energy between a particle and the surface is defined by the contact area between the particle and the adhesive domains. It is shown that the adsorption behaviour is controlled not only by the total area of the adhesive regions, but also by the geometry of the surface pattern. In particular, the domain size is found to have a significant effect on the adsorption efficiency. The most pronounced effect is observed when the particle and domain sizes are equal, which leads to enhanced adsorption at intermediate values of the chemical potential. At high values of the chemical potential, however, when the particle surface coverage increases, steric effects become important, which weakens the influence of the  surface pattern geometry. The obtained results demonstrate that the adsorption efficiency and surface organization of particles can be tuned by choosing the size, coverage, and spatial arrangement of adhesive domains. This study may be useful in the design of functional surfaces, selective adsorption platforms, biosensors, and affinity-based cell sorting systems.
\end{abstract}
	
\section{Introduction}

Systems of micro- and nanoscale particles interacting with chemically patterned surfaces represent an important class of interfacial problems in modern science and engineering. Such systems arise in a variety of processes, including selective adsorption, micro- and nano- structure formation, and the design of functional coatings and sensing elements~\cite{Maury2008,VanDommelen2018,Steinbach2013,Burkhardt2010,Xing2020,Beggiato2022,Jambhulkar2024,Rao2024}. Similar phenomena are also encountered in biophysical systems, where adsorption processes involve biomolecules and living cells interacting with patterned surfaces, particularly in cell sorting applications~\cite{didar2010adhesion,Badenhorst2025}.
A central question concerns how the surface pattern influences the spatial organization and adsorption behaviour of particles and biological entities at the interface, in particular their lateral arrangement and correlations. Computer simulation techniques provide a systematic tool for addressing this problem by analyzing surface coverage together with the structural organization of adsorbed particles as a function of the parameters of the surface pattern.

The effect of surface patterning on adsorption has been extensively studied in polymeric systems, providing important conceptual insights that are also directly relevant to the adsorption of particles. For example, Monte Carlo (MC) simulations by Semler and Genzer~\cite{SemlerGenzer2003,SemlerGenzer2004} demonstrated that the size of chemically different surface domains plays a crucial role in copolymer adsorption on patterned substrates. Their results showed that efficient selective adsorption and ordering occur only when the domain size is comparable to or larger than the characteristic length of polymer blocks, enabling molecular recognition and alignment with the surface pattern. For smaller domains, polymers are unable to adapt to varying surface properties, leading to reduced adsorption and disordered configurations. Related theoretical work by Chervanyov and Heinrich~\cite{ChervanyovHeinrich2006} further showed that random (stochastic) chemical heterogeneity can enhance polymer adsorption more effectively than regular patterning with the same average interaction strength due to the presence of multiple favourable local adsorption sites.

Besides polymer adsorption, simpler molecular systems have also been investigated to elucidate the role of surface heterogeneity. MC studies of a square lattice model for dimer adsorption on random heterogeneous surfaces~\cite{Nitta1997} demonstrated that not only the concentration but also the spatial distribution of adhesive sites strongly affects surface coverage due to blocking and packing constraints. 
The kinetics of particles deposition on heterogeneous surfaces has been extensively investigated within the random sequential adsorption (RSA) model, particularly for surfaces with periodically arranged adhesive regions of a square shape~\cite{StojiljkovicVrhovac2017}. This study demonstrate that the geometry of the underlying pattern, through the interplay between a size of adhesive regions and spacing between them, controls both the approach to the jamming limit and the structural organization of the resulting monolayer.

The adsorption of colloidal particles and macromolecules on solid surfaces is often governed by irreversible deposition processes, in which adsorbed particles neither desorb nor diffuse along the surface. Such systems cannot be described within equilibrium statistical mechanics and instead exhibit history-dependent, non-equilibrium behaviour controlled by excluded-volume interactions~\cite{Senger2000}. A central concept in this context is the progressive reduction of the available surface for adsorption, which governs the kinetics and leads to a jammed state characterized by slow, typically power-law, approach to saturation~\cite{Schaaf2000}. This process is commonly described by the random sequential adsorption (RSA) model.
Based on this approach, studies have shown that surface heterogeneity can qualitatively modify adsorption behaviour. In RSA-type models, deposition kinetics is highly sensitive to the spatial definition of adsorption sites: even very small deviations from a perfectly ordered lattice restore continuum-like behaviour, resulting in a crossover from exponential to power-law approach to jamming~\cite{PrivmanYan2016}. More generally, studies of patterned substrates demonstrate that the geometry and spacing of active regions control both the late-time kinetics and the structure of the adsorbed layer, enabling transitions between lattice-dominated and continuum regimes~\cite{StojiljkovicVrhovac2017}.
Equilibrium descriptions, such as the random site model (RSM), further show that quenched disorder in the spatial distribution of adsorption sites leads to non-trivial adsorption isotherms and coverage limits that cannot be reduced to homogeneous-surface behaviour. In these systems, steric exclusion combined with random site placement introduces correlations that significantly modify thermodynamic properties, with well-defined mappings to continuum adsorption only in the limit of high site density~\cite{TalbotTarjusViot2008}. Numerical and analytical results indicate that these effects are most pronounced when the characteristic length scale of disorder is comparable to the particle size~\cite{OleyarTalbot2007}.

The adsorption and deposition of particles on regular substrate patterns have also been studied within the RSA model in~\cite{CadilheAraujoPrivman2007}. This simulation work established that surface patterning strongly affects the morphology, packing efficiency, and saturation properties of particle monolayers. In particular, it was demonstrated that pre-patterned substrates significantly modify the structural arrangement of deposited particles, leading to distinct morphological regimes depending on the particle size and the characteristics of the square lattice. Subsequent investigations extended these ideas to nanopatterned substrates, revealing that surface patterning controls spatial correlations and domain formation in a particle monolayer~\cite{AraujoCadilhePrivman2008}, as well as the irreversible adsorption behaviour of polydisperse particles~\cite{Marques2012}. Related geometrical effects in two-dimensional adsorption were also analyzed in~\cite{Lebovka2023PRE}, who studied two-stage random sequential adsorption of discorectangles and disks on a planar surface and showed how particle shape and adsorption sequence influence the structure and final coverage of the adsorbed layer.

Beyond idealized particle models, patterned and structured surfaces play a crucial role in controlling adsorption processes in nanotechnology and biointerfaces. Large-area nanopatterning techniques enable precise control over nanoparticle and biomolecule arrangement~\cite{Barad2021}, while micro- and nano-structured surfaces have been shown to regulate protein adsorption and cell attachment~\cite{Khalili2015,Cai2020}. Biomimetic surface designs, such as honeycomb patterned or mushroom-shaped microstructured surfaces, can enhance or suppress adhesion through purely geometric effects~\cite{Carbone2011,Chen2015}. 
Experimental studies further demonstrate that surface topography and mechanical cues at the micro- and nanoscale critically influence cell adhesion, morphology, and mechanosensing~\cite{Ghassemi2012,PolacheckChen2016,Zhang2022}, 
with direct implications for antifouling, biomedical coatings, and tissue engineering applications~\cite{LiGuo2019,Uesugi2022,EskhanJohnson2022,XuSiedlecki2017}.

A number of experimental studies reported in the literature provide valuable insight into particle adsorption and deposition on heterogeneous and structured surfaces. 
In particular, the work by Rizwan and Bhattacharjee~\cite{RizwanBhattacharjee2009} investigated how surface charge heterogeneity affects the deposition of colloidal particles. 
The authors fabricated substrates with alternating positively and negatively charged stripes using self-assembled monolayers of alkanethiols terminated with carboxyl and amine groups. 
Deposition experiments were performed under quiescent conditions using sulphated polystyrene microspheres and fluorescent polystyrene nanoparticles. 
The results showed that particles preferentially deposit near the edges of favourable (positively charged) stripes. 
This effect was found to depend on both the width of unfavourable stripes and the particle size: narrower unfavourable regions or smaller particles lead to a less pronounced edge accumulation. 
Monte Carlo simulations based on a random sequential adsorption (RSA) model reproduced the experimental observations, demonstrating that even a simple binary surface model can adequately capture the essential features of the deposition morphology.

Nanostructured surfaces are also of considerable interest for applications in tissue engineering, regenerative medicine, and biosensing, where controlled cell attachment and detachment are required. 
In the study by Yu, Johnson, and L{\'o}pez~\cite{YuJohnsonLopez2014}, an innovative approach was proposed for fabricating structured thermoresponsive surfaces enabling controlled detachment of anchorage-dependent cells. 
The authors employed polymer brushes of poly(N-isopropylacrylamide) (PNIPAAm), which exhibit a lower critical solution temperature (LCST) of approximately 32~$^\circ$C. 
At temperatures above the LCST, PNIPAAm becomes hydrophobic, promoting cell adhesion and proliferation, whereas cooling below the LCST leads to spontaneous cell detachment. 
Importantly, nanopatterning was shown to overcome limitations associated with the critical polymer thickness required for efficient cell detachment on flat surfaces and enabled preferential cell adhesion to the nanopatterned regions.

Another experimental study that motivated the present work is that by Kumar, Parajuli, and Hahm~\cite{KumarParajuliHahm2007}, who introduced a method for fabricating high-density, two-dimensionally ordered protein nanoarrays using templates based on amphiphilic diblock copolymers. 
Using thin films of PS-\emph{b}-P4VP, the authors achieved regular hexagonal arrangements of proteins. 
A key mechanism underlying this approach is the selective adsorption of proteins onto adhesive domains formed through microphase separation of the diblock copolymers. 
Specifically, the hydrophilic P4VP domains act as adhesive sites that selectively bind proteins, while the hydrophobic PS matrix ensures spatial separation, enabling nanoscale control over protein positioning and the formation of functional protein nanoarrays.

Despite these extensive studies, the equilibrium adsorption of hard particles on geometrically patterned adhesive domains, particularly the role of domain size in shaping adsorption isotherms and inducing qualitative changes in adsorption behaviour, remains comparatively unexplored. This gap motivates the present study, which focuses on the interplay between particle and adhesive domain sizes, the surface density of domains and the equilibrium adsorption of particles on patterned surfaces.

In this work, we consider a two-dimensional model of particles (disks) adsorbing onto a surface structured as regular and disordered square circular adhesive domains. The particles interact with the domains via an attractive potential proportional to the area of their contact (overlap) with the adhesive regions and are free to move within the plane of the surface. At the same time, particle-particle interactions are modelled as hard-disk repulsions.
The aim of this study is to investigate the adsorption efficiency and structural arrangement of particles on a patterned surface depending on the size of the adhesive domains and their surface density. Using Monte Carlo simulations in the grand canonical ($\mu$VT) ensemble we obtain adsorption isotherms and radial distribution functions, drawing the conclusions about the influence of surface geometry on the adsorption characteristics of particles for different domain sizes and relate these effects to the structural properties of the adsorbed particle layer.

\section{Model description}

\begin{figure}[!htb]
	\centering
	\includegraphics[width=0.35\linewidth]{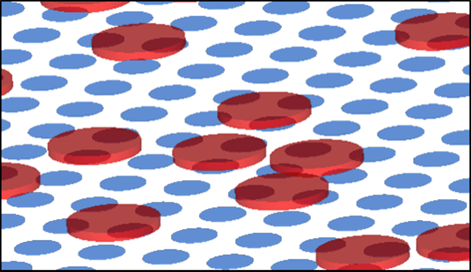}
	\caption{Illustration of the relevant three-dimensional model system. Particles (red disks) interact with attractive domains (blue disks) arranged 
		in a square lattice. The transparency of the red particles is used to illustrate the contact interface between particles and adhesive domains.}
	\label{fig:illustration}
\end{figure}

The model presented in this work is based on a three-dimensional model for two-component system of adsorbing particles on the surface patterned with adhesive 
domains, which was originally introduced in~\cite{Badenhorst2025}. 
We adapt this model to a two-dimensional description of a one-component system, retaining the essential geometric and interaction features relevant to particle adsorption on patterned surfaces. Despite the reduced dimensionality, the model still captures equilibrium adsorption of disk-like particles from a dilute bulk phase onto the patterned surface, leading to the formation of a particle monolayer. The corresponding three-dimensional representation is shown in~Fig.~\ref{fig:illustration}.

\begin{figure}[!htb]
	\centering
	\includegraphics[width=0.3\linewidth]{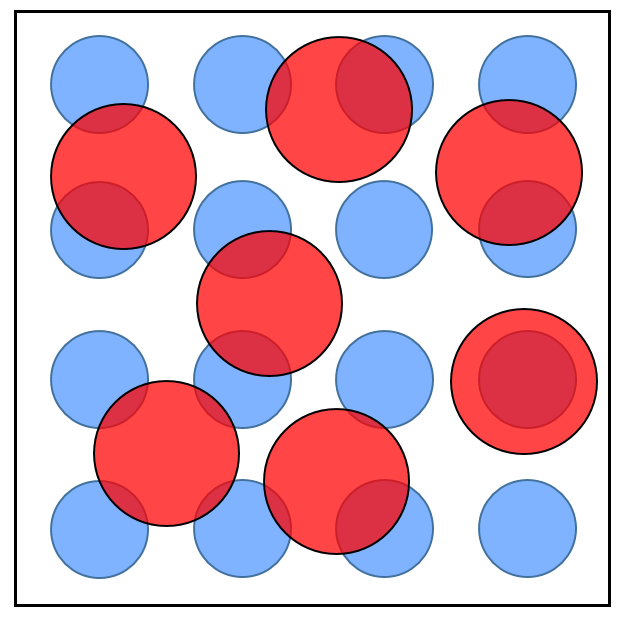} \qquad
	\includegraphics[width=0.3\linewidth]{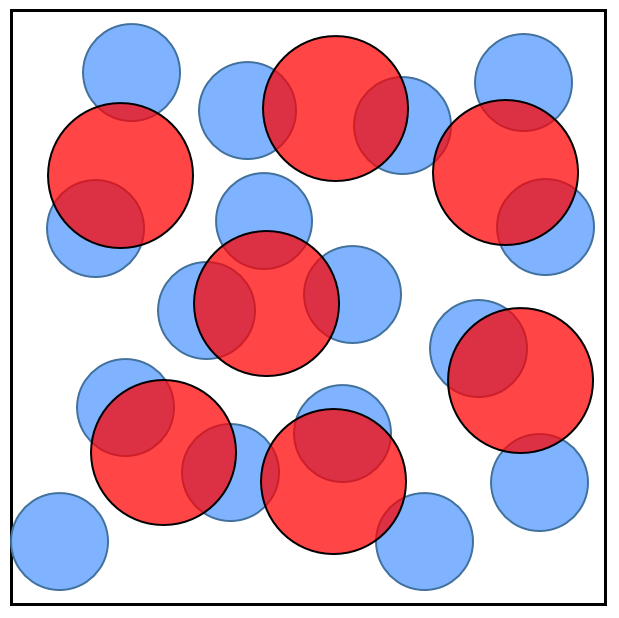}	
	\caption{A two-dimensional model of a patterned surface with adhesive domains and adsorbed particles. Particles (red disks) interact with attractive domains (blue disks) arranged either in a square lattice (left panel) or in a random configuration (right panel). The colour notation is the same as in Fig.~\ref{fig:illustration}.}.
	\label{fig:domains}
\end{figure}

Thus, the model considers a two-dimensional plane on which adhesive domains are arranged either in an ordered or disordered structure. These domains are treated as fixed regions on the surface that interact attractively with particles adsorbed onto them.
Each domain has a circular shape with a fixed diameter $D_\mathrm{d}=2R_\mathrm{d}$.
In one case, the domains are positioned at the nodes of a square lattice (ordered arrangement) with the lattice parameter $a$ determining the distance between 
the centers of two neighbouring domains (in the $X$ and $Y$ directions).
As a result, the $N_\mathrm{d}=N_x \times N_y$ domains forms a regular pattern.
In another case, the disordered pattern is formed by $N_\mathrm{d}$ domains distributed randomly along the surface without overlapping with each other.
In both cases, the positions of domains are fixed, while the particles are modelled as hard disks of diameter $D_{\rm p}=2R_{\rm p}$ (see Fig.~\ref{fig:domains}), which can move freely in $XY$-plane, under condition that the distance between any two particles cannot be smaller than $D_{\rm p}$.
The number of particles, $N_\mathrm{p}$, is either fixed, corresponding to Monte Carlo simulations in the canonical ensemble (NVT), or allowed to fluctuate when the grand canonical ensemble ($\mu$VT) is employed. Periodic boundary conditions are applied in both spatial directions, $X$ and $Y$, which effectively eliminates boundary effects and allows an infinite surface to be modelled.

\begin{figure}[!htb]
	\centering
	\includegraphics[width=0.3\linewidth]{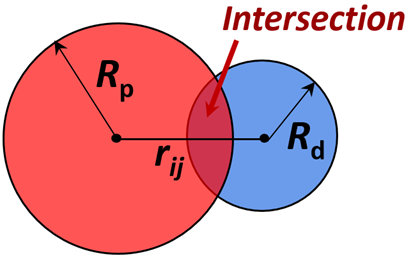}
	\caption{Intersection of two disks of radii $R_{\rm p}$ and $R_{\rm d}$ whose centers are separated by a distance $r_{ij}$. The colour notation is the same as in Fig.~\ref{fig:illustration}.}	
	\label{fig:intersection}
\end{figure}

The potential energy of interaction between particle $i$ with coordinates $(x_i,y_i)$, and domains is defined as the sum of pair interaction energies between this particle and the domains with which it is in contact (overlaps). 
The pair interaction energy between a particle and a domain is assumed to be proportional to the contact area between the particle and the domain, which is equal to the area of overlap between the particle and the domain, $s_{\rm pd}(r_{ij})$, and depends on the distance between their centers, $r_{ij}$. 
Accordingly, the potential energy of particle $i$ can be written as
\begin{equation}
	U_{s\rm p}^{\mathrm{ads}}(x_i,y_i)
	=
	\sum_{\{j \, | \, r_{ij} \le r_{\rm pd}\}}
	A_{\rm pd}\, s_{\rm pd}(r_{ij}),
	\label{eq:Up_ads}
\end{equation}
where $r_{\rm pd}=R_{\rm p}+R_{\rm d}$ is the interaction range between a particle and a domain, and $A_{\rm pd}$ is the attraction strength parameter between the particle and the domain. 
The sum in Eq.~\eqref{eq:Up_ads} runs over all domains $j$ located within a distance $r_{\rm pd}$ from particle $i$.
The overlap area of two disks with radii $R_{\rm p}$ and $R_{\rm d}$, whose centers are separated by a distance
\begin{equation}
	r_{ij}=\sqrt{(x_i-x_j)^2+(y_i-y_j)^2},
\end{equation}
is given by
\begin{equation}
	s_{\rm pd}(r_{ij})=
	\begin{cases}
		s_{\rm d}, & r_{ij}\le |R_{\rm p}-R_{\rm d}|,\\[6pt]
		s_{\mathrm{int}}(r_{ij}), & |R_{\rm p}-R_{\rm d}|< r_{ij}\le R_{\rm p}+R_{\rm d},\\[6pt]
		0, & r_{ij}>R_{\rm p}+R_{\rm d},
	\end{cases}
	\label{eq:overlap_piecewise}
\end{equation}
where $s_{\rm d}$ denotes the area of the smaller disk, and the intersection area $s_{\mathrm{int}}(r_{ij})$ (see Fig.~\ref{fig:intersection}) is given by
\begin{equation}
	\begin{aligned}
		&s_{\mathrm{int}}(r_{ij}) =
		R_{\rm p}^{2}\arccos\!\left(\frac{r_{ij}^{2}+R_{\rm p}^{2}-R_{\rm d}^{2}}{2R_{\rm p}r_{ij}}\right)\\
		&+
		R_{\rm d}^{2}\arccos\!\left(\frac{r_{ij}^{2}+R_{\rm d}^{2}-R_{\rm p}^{2}}{2R_{\rm d}r_{ij}}\right)\\
		&
		-\frac{1}{2}\sqrt{\left[(R_{\rm p}+R_{\rm d})^2-r_{ij}^2\right]
			\left[r_{ij}^2-(R_{\rm d}-R_{\rm p})^2\right]} .
	\end{aligned}
	\label{eq:intersection_area}
\end{equation}

This approach enables the description of particle adsorption on a surface mediated by adhesive domains, with the adhesion energy depending directly on the contact area between the particles and the domains. It thus captures the geometric characteristics of both the patterned surface and the particles adsorbed on it. The dependence of the potential energy on the overlap area makes it possible to model domain-induced particle attachment to the surface, that is, the ability of adhesive domains to retain particles in their vicinity. As the overlap area increases, the particle energy decreases, thereby increasing the probability of finding a particle within the domain region. Particles therefore tend to be distributed over the surface so as to maximize their contact with the domains. This behaviour, however, depends on the domain size, surface coverage, and spatial distribution of the domains, i.e. on the geometry of the surface pattern. In what follows, we consider several parameter sets to elucidate the influence of these factors.

\section{Results and discussion}

\subsection{Computer simulation details}

A series of Monte Carlo simulations was carried out for hard disks adsorbed on a surface with adhesive domains, exploring a range of system parameters defining the surface pattern geometry. In all cases, the interaction parameter $A_{\rm dp}$, which describes the strength of the attractive disk-domain interaction, was fixed at a relatively high value of $A_{\rm{dp}} = \beta A_0 \left(\frac{\pi D_{\rm p}^2}{4}\right)^{-1} = -12.732$.
This value corresponds to an interaction energy, which is ten times larger than the thermal energy $k_{\rm B} T = \beta^{-1}$ when a particle of diameter $D_{\rm p}$ interacts with a domain over its entire area $\pi D_{\rm p}^2/4$.
Only the domain diameter $D_{\rm d}$ and the surface coverage fraction of the domains $\sigma_{\rm d}$ were varied in our study.
The surface coverage fraction of domains is defined as
\begin{equation}
	\sigma_{\rm d} = \frac{\pi}{4} D_{\rm d}^2 \rho_{\rm d}, 
	\qquad
	\rho_{\rm d} = \frac{N_{\rm d}}{S},
	\qquad
	S = L_x \times L_y,
\end{equation}
where $\sigma_{\rm d}$ is equivalent to the surface packing fraction of the domains, $\rho_{\rm d}$ is the surface number density of domains, 
$N_{\rm d}$ is the total number of domains, and $S$ is the total surface area corresponding to the area of the two-dimensional simulation box. 
For a meaningful comparison of the results, each series of simulations was carried out by fixing one of these parameters. Specifically, when examining the effect of domain size, the number of domains, i.e. their surface density, was adjusted so that the surface coverage fraction $\sigma_{\rm d}$ remained constant. In contrast, when analyzing the effect of the surface coverage fraction, the domain size $D_{\rm d}$ was kept fixed, while the surface number density of domains, $\rho_{\rm d}$, was varied.
In the case of ordered domains (square lattice), the total number of domains is given by $N_{\rm d} = n_x \times n_y$, where $n_x$ and $n_y$ denote the number of lattice nodes along the $X$ and $Y$ axes, respectively. The lattice parameter (i.e., the spacing between lattice nodes) is then given by
$a = L/n$, here $L = L_x = L_y$ and $n = n_x = n_y$. For the case of disordered domains we choose the same values of $N_{\rm d}$.
In this study, we explore domain packing fractions in the range $\sigma_{\rm d} = 0.256$--$0.785$, while the domain sizes were chosen as $D_{\rm d}/D_{\rm p} = 0.5$, $1.0$, $1.5$, and $2.0$. Thus, we consider scenarios in which the domains are either half the size of the particles, equal in size, or $1.5$ and $2.0$ times larger.

First, we examine how the adsorption isotherms vary with domain size at fixed domain surface coverage, $\sigma_{\rm d}$.
Clearly, varying the domain size modifies the geometry of the adhesive regions, which inevitably affects particle adsorption.
However, the magnitude of this effect and the direction of the resulting shift in the adsorption isotherms are not obvious \emph{a priori}. To address this question, we performed Monte Carlo simulations in the grand canonical ensemble ($\mu$VT) at a fixed temperature, varying the chemical potential over the range $\mu^*=\beta\mu=-12.0$ to $12.0$. Depending on the chemical potential, the system reached an equilibrium particle density corresponding to the imposed thermodynamic conditions. Each simulation consisted of $500{,}000$ Monte Carlo steps to reach equilibrium, followed by an additional $500{,}000$ steps for data collection. 
During the production stage, the particle density was recorded every ten Monte Carlo steps, from which the average value and its statistical errors were calculated.

\subsection{Adsorption isotherms}

\begin{figure}[!htb]
	\centering
	\includegraphics[width=0.49\linewidth]{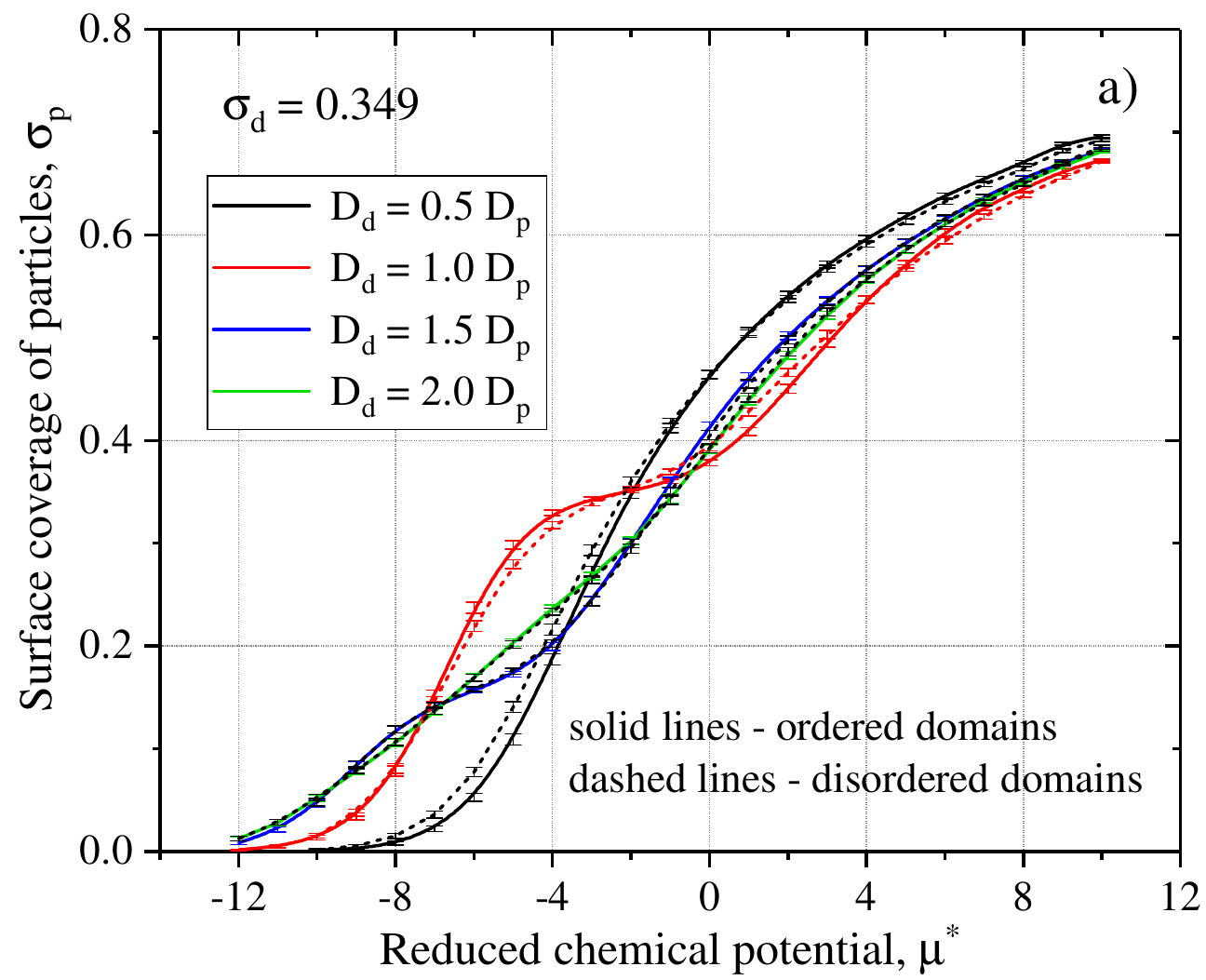}
	\includegraphics[width=0.49\linewidth]{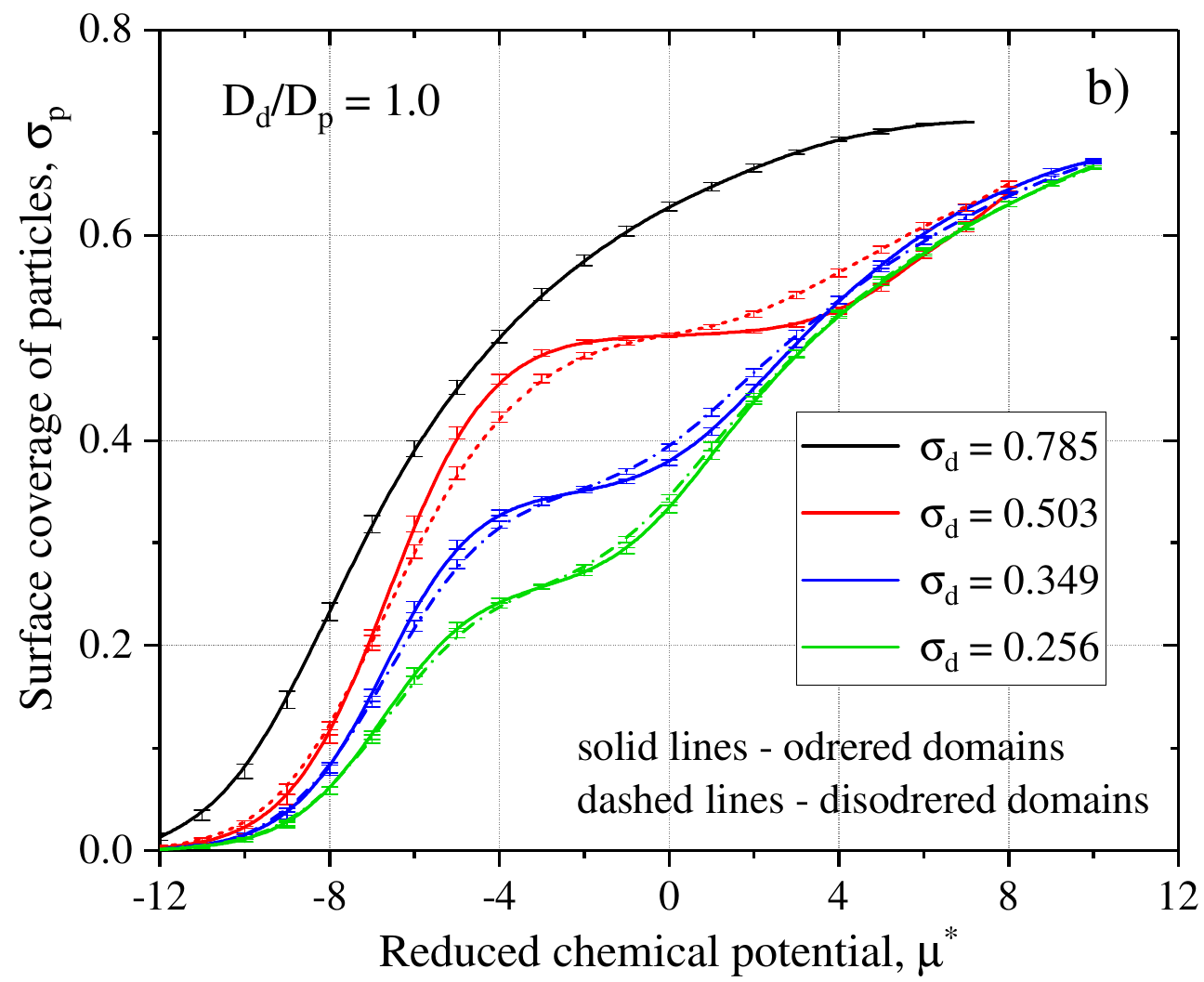}	
	\caption{Adsorption isotherms of disk-like particles on a patterned adhesive surface: (a)~the domain size is $D_{\rm d}/D_{\rm p}=0.5$, $1.0$, $1.5$, or $2.0$, while the domain surface coverage is fixed at $\sigma_{\rm d}=0.349$; (b)~the domain surface coverage is $\sigma_{\rm d}=0.256$, $0.349$, $0.503$, or $0.785$, while the domain size is fixed at $D_{\rm d}/D_{\rm p}= 1.0$.}
	\label{fig:isotherms}
\end{figure}

The calculated particle surface coverage, $\sigma_{\rm p}$, as a function of the reduced chemical potential, $\mu^*$, is presented in Fig.~\ref{fig:isotherms} for patterned adhesive surfaces with ordered (solid lines) and disordered (dashed lines) domain arrangements. In both panels, the adsorption isotherms increase monotonically with $\mu^*$ in the range from $-12.0$ to $12.0$, as expected for equilibrium adsorption in the grand canonical ensemble. At the same time, their shape and relative position depend strongly on the domain size, the domain surface coverage, and, to a lesser extent, the spatial arrangement of the domains.

The effect of domain size at fixed domain surface coverage, $\sigma_{\rm d}=0.349$, is presented in Fig.~\ref{fig:isotherms}a. Four values of the size ratio are considered: $D_{\rm d}/D_{\rm p}=0.5$, $1.0$, $1.5$, and $2.0$. For small domains, $D_{\rm d}/D_{\rm p}=0.5$, the adsorption isotherm is smooth over the whole range of chemical potentials. This behaviour is similar to that expected for a weakly heterogeneous attractive surface. Since the domains are two times smaller than the particles, a particle cannot fully overlap with a single domain. Instead, it interacts with a locally averaged attractive pattern formed by several smaller adhesive domains. As a result, the detailed geometry of the surface pattern has only a weak influence on the adsorption process.

A qualitatively different behaviour is observed for $D_{\rm d}/D_{\rm p}=1.0$, when the domain size is equal to the particle size. In this case, the adsorption isotherm exhibits a pronounced change in slope. This indicates that adsorption no longer proceeds as a simple smooth uptake process. At low and intermediate chemical potentials, particles preferentially occupy positions where they can almost completely overlap with individual adhesive domains. Therefore, the system with $D_{\rm d}/D_{\rm p}=1.0$ shows enhanced adsorption compared with the case $D_{\rm d}/D_{\rm p}=0.5$. This enhancement is especially visible up to approximately $\mu^*\simeq -1.97$, where the isotherms for $D_{\rm d}/D_{\rm p}=0.5$ and $D_{\rm d}/D_{\rm p}=1.0$ intersect.
This crossover has a clear geometrical interpretation. At the intersection point, the particle number density is approximately $\rho_{\rm p}=0.444$. This value coincides with the domain number density for the system with $D_{\rm d}/D_{\rm p}=1.0$ and $\sigma_{\rm d}=0.349$, which corresponds approximately to one domain per particle. Below this point, the one-particle--one-domain matching is favourable: a single particle can almost completely occupy one adhesive domain and thus gain the maximum attractive contribution available from that domain. Above this point, the most favourable adsorption positions are already occupied, and additional particles must either share adhesive regions with neighbouring particles or occupy less favourable positions between domains. Consequently, the average attractive energy gained per particle decreases, while steric and packing constraints become increasingly important.
This explains why, at higher chemical potentials, the tendency reverses in favour of smaller domains. At fixed $\sigma_{\rm d}=0.349$, smaller domains are more numerous and more uniformly distributed over the surface. Therefore, at high particle coverages, a larger number of particles can still maintain partial contact with adhesive regions. In contrast, for $D_{\rm d}/D_{\rm p}=1.0$, the number of favourable one-domain--one-particle positions is limited, and once these positions are filled, further adsorption becomes less energetically efficient.

For larger domains, $D_{\rm d}/D_{\rm p}=1.5$ and $2.0$, particles can also be accommodated within adhesive regions. However, because the total domain surface coverage is fixed at $\sigma_{\rm d}=0.349$, increasing the domain size necessarily reduces the number of domains. Thus, larger domains provide more extended adhesive regions, but fewer such regions are available on the surface. As the number of adsorbed particles increases, the favourable space on these domains is exhausted more rapidly. This explains why, in the approximate range $\mu^*\simeq -8.0$ to $-1.0$, the isotherms for larger domains lie below the isotherm for $D_{\rm d}/D_{\rm p}=1.0$. Nevertheless, they remain more favourable than the case of small domains, $D_{\rm d}/D_{\rm p}=0.5$, up to approximately $\mu^*\simeq -4.0$. At higher chemical potentials, the advantage of larger domains gradually disappears because adsorption becomes increasingly controlled by packing constraints rather than by the maximum overlap with individual domains.

The influence of domain surface coverage at fixed domain size is analyzed in Fig.~\ref{fig:isotherms}b. Here the domain size is fixed at $D_{\rm d}/D_{\rm p}=1.0$, since the most pronounced effect of surface pattern heterogeneity is observed when the domains and particles have the same size. Four values of the domain surface coverage are considered: $\sigma_{\rm d}=0.256$, $0.349$, $0.503$, and $0.785$. As expected, increasing $\sigma_{\rm d}$ shifts the adsorption isotherms towards lower chemical potentials. This means that a given particle surface coverage, $\sigma_{\rm p}$, can be reached at a lower $\mu^*$ when the total attractive area on the surface is larger.
This effect is particularly strong at low and intermediate chemical potentials. For the largest domain surface coverage, $\sigma_{\rm d}=0.785$, which corresponds to the maximum packing fraction of disks arranged on a square lattice, the particle surface coverage starts to increase at much lower values of $\mu^*$ than for $\sigma_{\rm d}=0.256$ or $0.349$. This reflects the larger number of favourable adsorption positions available on the surface. At lower domain coverages, the adhesive domains are less numerous and more separated. Therefore, adsorption proceeds in a more clearly two-stage manner: first, particles occupy the most favourable positions where they can strongly overlap with individual domains; then, after these positions are filled, further adsorption occurs more slowly because particles must occupy less favourable regions and the role of steric constraints increases. 
This two-stage character is most clearly manifested for $\sigma_{\rm d}=0.503$, where a plateau-like intermediate region appears in the range $\mu^*\simeq -1.0$ to $1.0$. In general, with increasing $\sigma_{\rm d}$ from $0.256$ to $0.785$, the total attractive area becomes larger and the adhesive regions are more widely available over the surface. As a result, the surface behaves more like a uniformly attractive substrate, and the adsorption isotherm becomes smoother. The distinction between favourable and less favourable adsorption positions is then reduced, because particles can more easily find positions with substantial overlap with adhesive domains.

The comparison between ordered and disordered domain arrangements in Fig.~\ref{fig:isotherms} shows that the overall adsorption behaviour is controlled mainly by $D_{\rm d}/D_{\rm p}$ and $\sigma_{\rm d}$. The difference between the corresponding solid and dashed curves is relatively small, especially at high chemical potentials, where the surface is already densely populated and adsorption is governed primarily by packing constraints. Nevertheless, some deviations between ordered and disordered patterns are visible at low and intermediate $\mu^*$. These deviations arise because a disordered surface contains a broader distribution of local environments: in some regions, domains are locally closer together, whereas in others they are farther apart than in the ordered square lattice. Such local variations tend to smear out the adsorption features associated with particle--domain size matching.

Overall, the results presented above demonstrate that adsorption on patterned adhesive surfaces is not determined solely by the total attractive area. At fixed $\sigma_{\rm d}=0.349$, changing the domain size from $D_{\rm d}/D_{\rm p}=0.5$ to $1.0$, $1.5$, and $2.0$ substantially changes the shape of the adsorption isotherms. The strongest deviation from smooth, homogeneous-surface-like behaviour occurs for $D_{\rm d}/D_{\rm p}=1.0$, where the domain and particle sizes match. This particle--domain matching enhances adsorption at low and intermediate chemical potentials, but becomes less efficient after the favourable one-particle--one-domain configurations are exhausted. Similarly, at fixed $D_{\rm d}/D_{\rm p}=1.0$, increasing the domain surface coverage from $\sigma_{\rm d}=0.256$ to $0.785$ promotes adsorption and makes the surface effectively more attractive and more homogeneous.

\subsection{Spatial distribution analysis}

\begin{figure*}[!htb]
	\centering
	\includegraphics[width=0.85\linewidth]{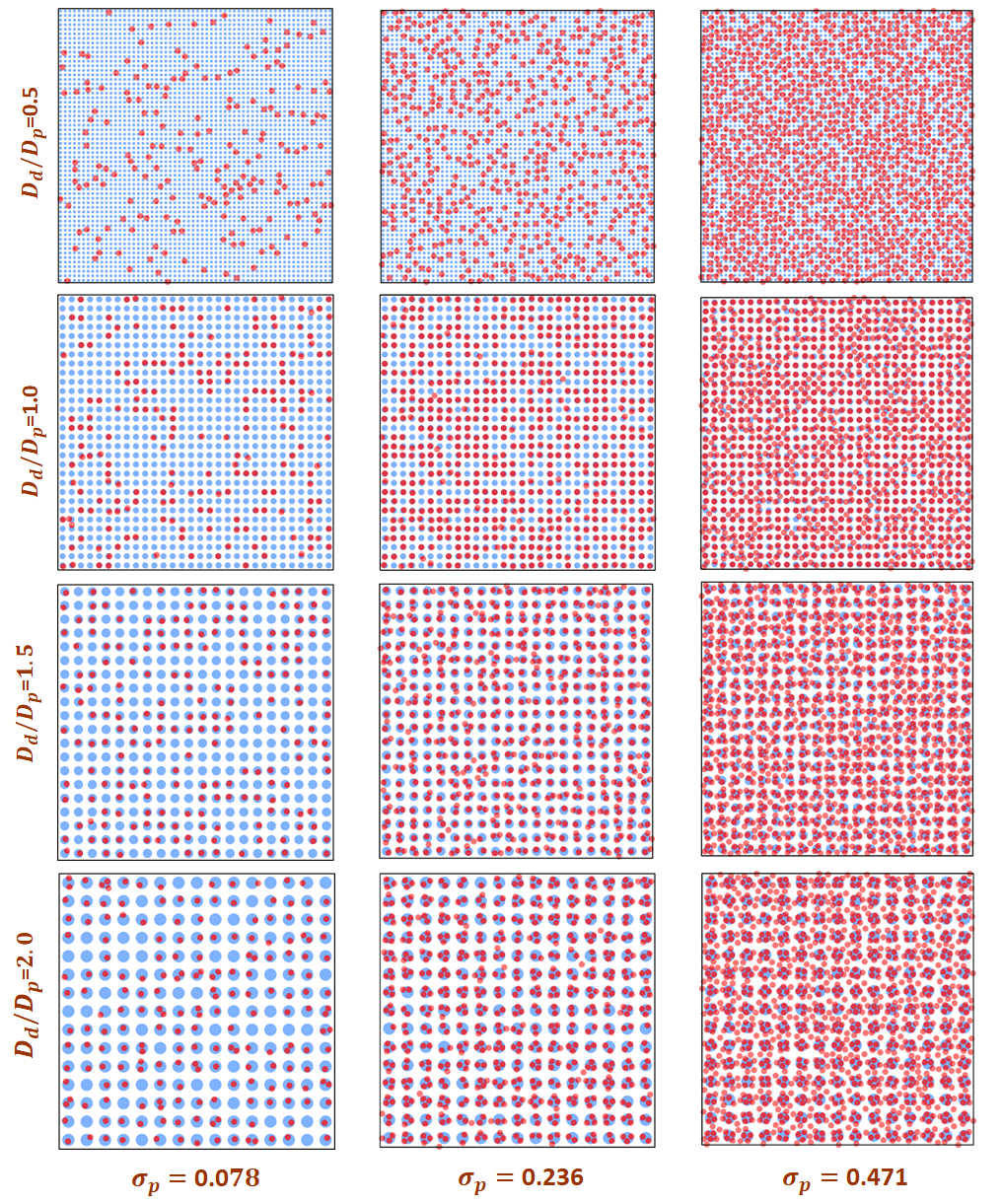}
	\caption{Configurations of particles adsorbed on a patterned surface with order domains of different sizes 
		$D_{\rm d}/D_{\rm p} = 0.5$, $1.0$, $1.5$, and $2.0$ (rows from top to bottom) and particle surface densities
		$\sigma_{\rm p} = 0.078$, $0.236$, and $0.471$ (columns from left to right) at $\sigma_{\rm d}=0.349$.}
	\label{fig:snapshots}
\end{figure*}

To further analyze the adsorption behaviour of particles on a patterned surface, we consider several representative configurations illustrating the lateral particle distribution at different particle coverages obtained from Monte Carlo simulation in the NVT ensemble. The snapshots in Fig.~\ref{fig:snapshots} show particle configurations on ordered patterned surfaces with different domain sizes, $D_{\rm d}/D_{\rm p}=0.5$, $1.0$, $1.5$, and $2.0$, at fixed domain surface coverage, $\sigma_{\rm d}=0.349$. The columns correspond to particle surface coverages $\sigma_{\rm p}=0.078$, $0.236$, and $0.471$, which represent low, intermediate, and relatively high adsorption regimes, respectively. These configurations therefore provide a direct visual illustration of the adsorption mechanisms suggested by the isotherms in Fig.~\ref{fig:isotherms}.

At the lowest particle coverage, $\sigma_{\rm p}=0.078$, the particles are well separated and steric constraints between them are weak. The particle distribution is therefore governed mainly by the particle--domain attraction. For small domains, $D_{\rm d}/D_{\rm p}=0.5$, the particles are distributed rather uniformly over the surface, because each particle overlaps with several small adhesive regions and experiences an effectively averaged attraction. In contrast, for $D_{\rm d}/D_{\rm p}=1.0$, the particles preferentially occupy positions close to individual domains. This reflects the favourable geometrical matching between the particle and domain sizes. For larger domains, $D_{\rm d}/D_{\rm p}=1.5$ and $2.0$, particles can also be accommodated within adhesive regions, but the number of such regions is smaller because the total domain surface coverage is fixed.

At the intermediate coverage, $\sigma_{\rm p}=0.236$, the effect of domain size becomes more pronounced. For $D_{\rm d}/D_{\rm p}=0.5$, the particle distribution remains relatively homogeneous, although the underlying domain pattern still affects the local probability of adsorption. For $D_{\rm d}/D_{\rm p}=1.0$, many particles are located near individual domains, illustrating the one-particle--one-domain adsorption mechanism discussed above. This regime corresponds to the range of chemical potentials where the adsorption isotherm shows a pronounced change in slope. For larger domains, $D_{\rm d}/D_{\rm p}=1.5$ and $2.0$, several particles may be located within or near the same enlarged adhesive region. However, because the number of domains is reduced, the distribution becomes more spatially heterogeneous, with particle-rich regions near the domains and less populated regions between them.

At the highest particle coverage, $\sigma_{\rm p}=0.471$, the surface is densely populated and steric as well as packing constraints become important. In this regime, the influence of the attractive pattern is partly masked by the high particle concentration. Nevertheless, the domain size still affects the lateral organization of the adsorbed layer. For $D_{\rm d}/D_{\rm p}=0.5$, the particles form a relatively uniform dense distribution, consistent with the smooth adsorption isotherm observed for small domains. For $D_{\rm d}/D_{\rm p}=1.0$, most favourable domain-centered positions are already occupied, and additional particles must occupy less favourable interdomain regions or share attractive regions with neighbouring particles. This explains why the advantage of domains of the same size as particles is lost at higher chemical potentials. For $D_{\rm d}/D_{\rm p}=1.5$ and $2.0$, the particles remain influenced by the larger adhesive regions, but the limited number of such regions restricts the adsorption capacity associated with direct particle--domain overlap.

\begin{figure}[!htb]
	\centering
	\includegraphics[width=0.49\linewidth]{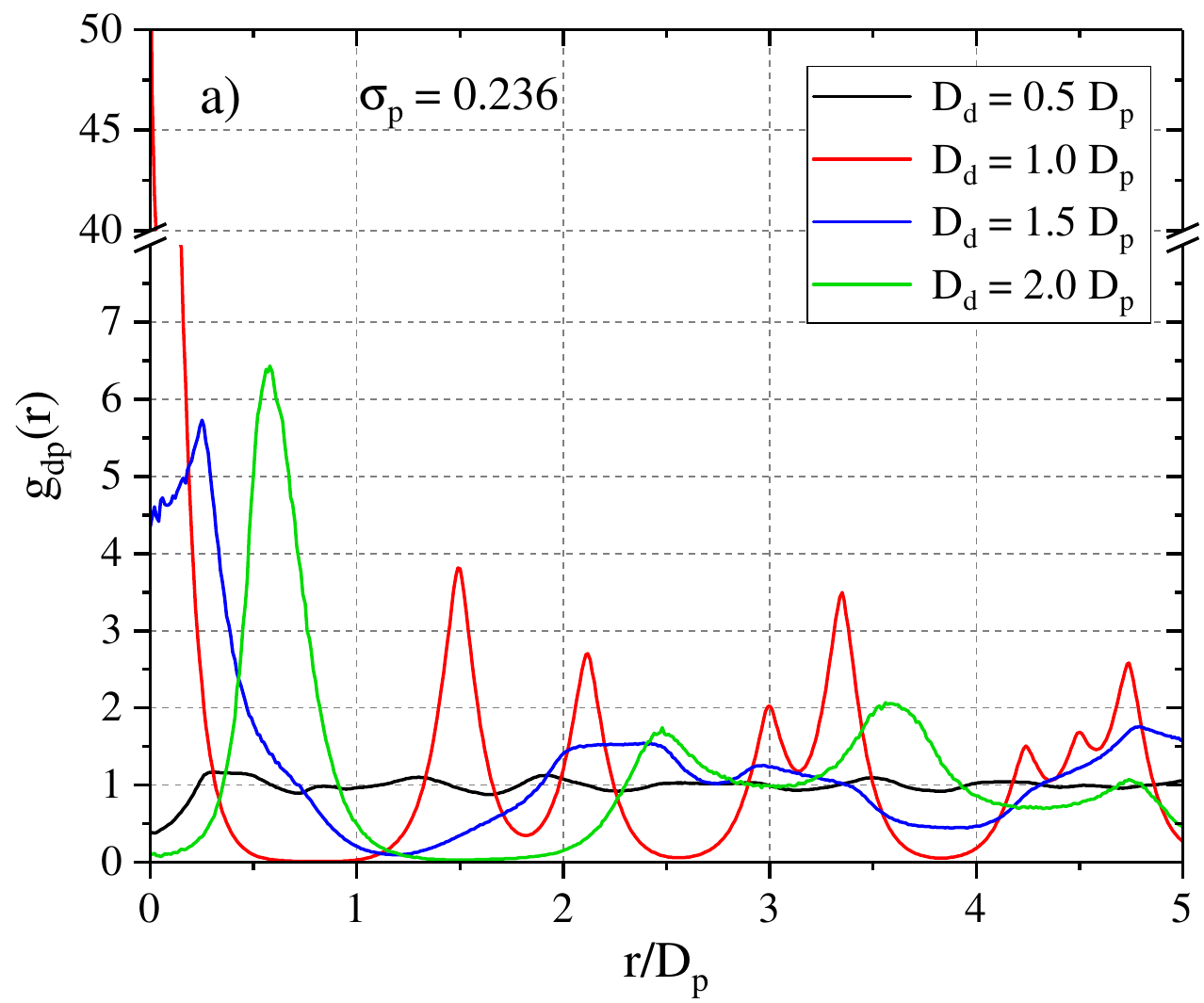}  
	\includegraphics[width=0.49\linewidth]{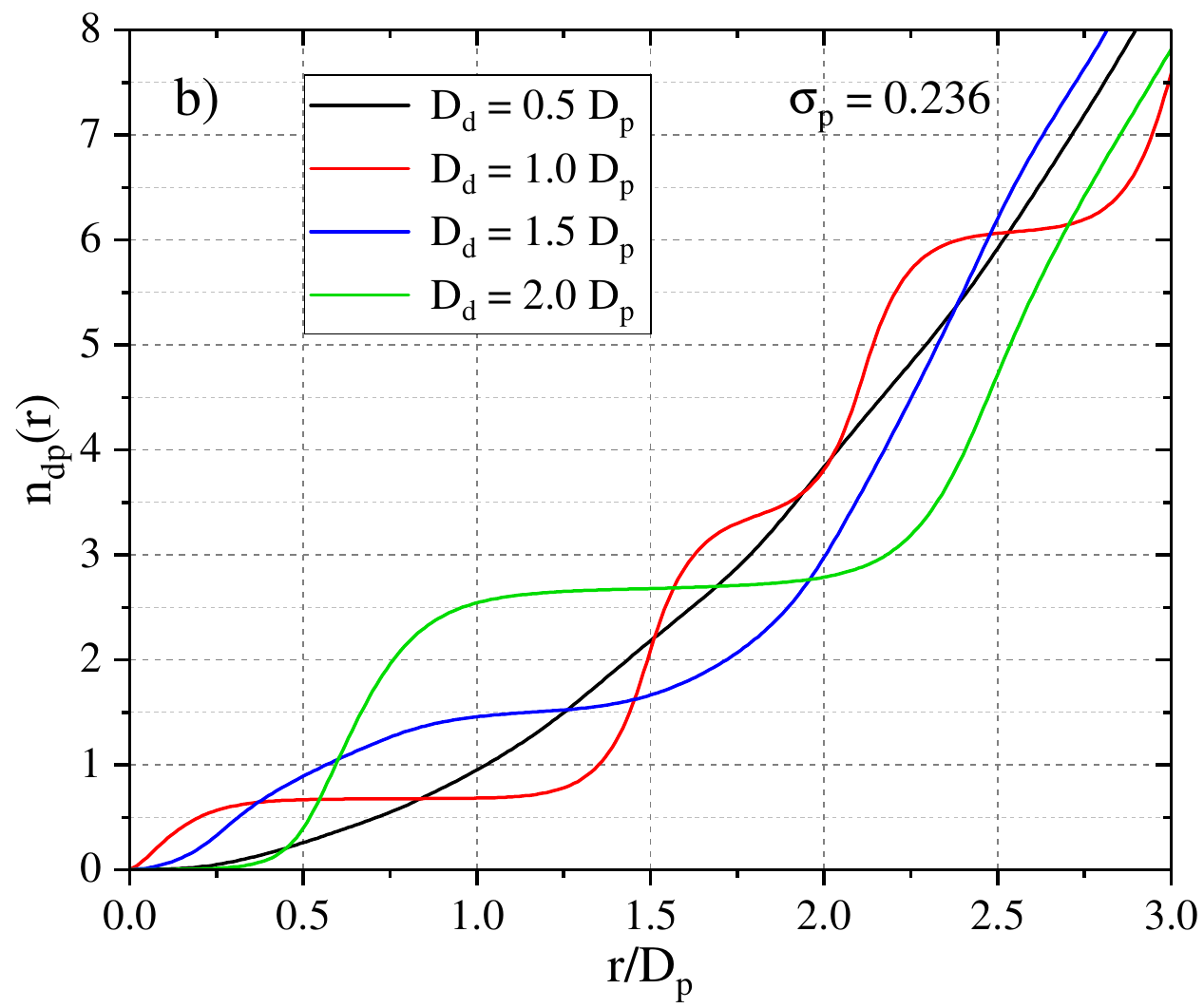}    
	\caption{Radial distribution function $g_{\rm dp}(r)$~(a) and cumulative coordination number $n_{\rm dp}(r)$~(b) for domain--particle pairs at different domain sizes $D_{\rm d}$ and fixed domain surface coverage $\sigma_{\rm d}=0.349$. The domains are arranged on a square lattice, and the particle surface coverage is $\sigma_{\rm p}=0.236$.}
	\label{fig:gdp}
\end{figure}

To complement the visual analysis of the particle configurations, we next consider the radial distribution functions and cumulative coordination numbers, which provide a more detailed description of the local structure of the adsorbed layer.
The average number of particles located around a given domain can be estimated from the cumulative coordination number $n_{\rm dp}(r)$. For this purpose, the radial distribution functions $g_{\rm dp}(r)$ were calculated and subsequently used to obtain the corresponding coordination numbers $n_{\rm dp}(r)$. Representative results for domain--particle correlations are shown in Fig.~\ref{fig:gdp} for ordered domain arrangements at fixed domain surface coverage $\sigma_{\rm d}=0.349$ and particle surface coverage $\sigma_{\rm p}=0.236$.
The radial distribution functions $g_{\rm dp}(r)$ in Fig.~\ref{fig:gdp}a clearly demonstrate that changing the domain size leads to qualitatively different spatial arrangements of particles with respect to the domain centers. This behaviour is governed by the tendency of particles to occupy energetically favourable positions on the surface. However, the location and the width of these favourable regions depend strongly on the domain size.

For small domains, $D_{\rm d}/D_{\rm p}=0.5$, the domain--particle correlation is relatively weak and broad. In this case, a single domain is much smaller than a particle and therefore cannot by itself provide a strongly localized adsorption site. As a result, the position of a particle is determined by the combined contribution of several nearby adhesive domains rather than by one particular domain. This explains why the corresponding first coordination shell is poorly defined.

A very different behaviour is observed for $D_{\rm d}/D_{\rm p}=1.0$. In this case, the domain and particle sizes are equal, and the most favourable configuration corresponds to almost complete overlap between a particle and a domain. This gives rise to a strong maximum of $g_{\rm dp}(r)$ at small distances from the domain center. The corresponding coordination number reaches approximately $n_{\rm dp}=0.68$ within the first coordination shell. This value is smaller than unity because not all domains are occupied at $\sigma_{\rm p}=0.236$, but it nevertheless indicates a clear tendency towards one-particle--one-domain adsorption.

For larger domains, $D_{\rm d}/D_{\rm p}=1.5$ and $2.0$, the first peak of $g_{\rm dp}(r)$ becomes broader and shifts to larger distances. This reflects the fact that a particle can be favourably located not only at the domain center but also at different positions within the extended adhesive region. Consequently, the attractive region associated with a single domain can accommodate more than one particle. This is confirmed by the cumulative coordination number in Fig.~\ref{fig:gdp}b: within the first coordination shell, $n_{\rm dp}$ increases from about $0.68$ for $D_{\rm d}/D_{\rm p}=1.0$ to approximately $1.52$ for $D_{\rm d}/D_{\rm p}=1.5$ and $2.68$ for $D_{\rm d}/D_{\rm p}=2.0$. These values should be regarded as average numbers of particles per domain. Locally, however, the actual number of particles associated with a particular domain may be either larger or smaller, as can also be seen from the snapshots in Fig.~\ref{fig:snapshots}.

The behaviour of $n_{\rm dp}(r)$ further illustrates the difference between particle-sized and larger domains. For $D_{\rm d}/D_{\rm p}=1.0$, the increase of $n_{\rm dp}(r)$ is rather step-like, indicating a relatively well-defined shell of particles located near individual domains. For $D_{\rm d}/D_{\rm p}=1.5$ and $2.0$, the increase is more gradual over a broader interval of distances, because particles can occupy a wider region around each domain center. Thus, increasing the domain size increases the number of particles that can be accommodated by a single adhesive domain, but at fixed $\sigma_{\rm d}$ it also reduces the total number of such domains on the surface.

\begin{figure}[!htb]
	\centering
	\includegraphics[width=0.49\linewidth]{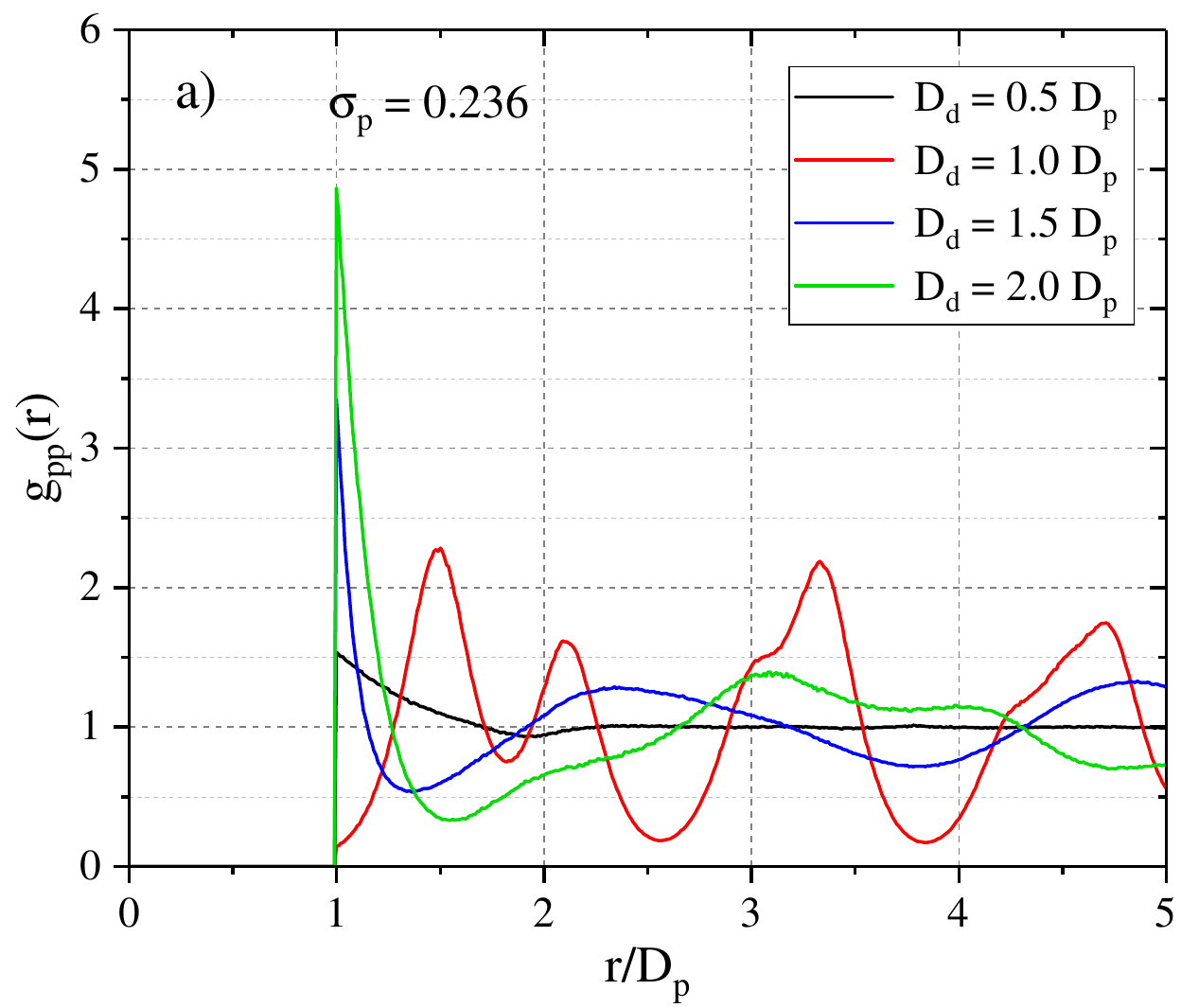} 
	\includegraphics[width=0.49\linewidth]{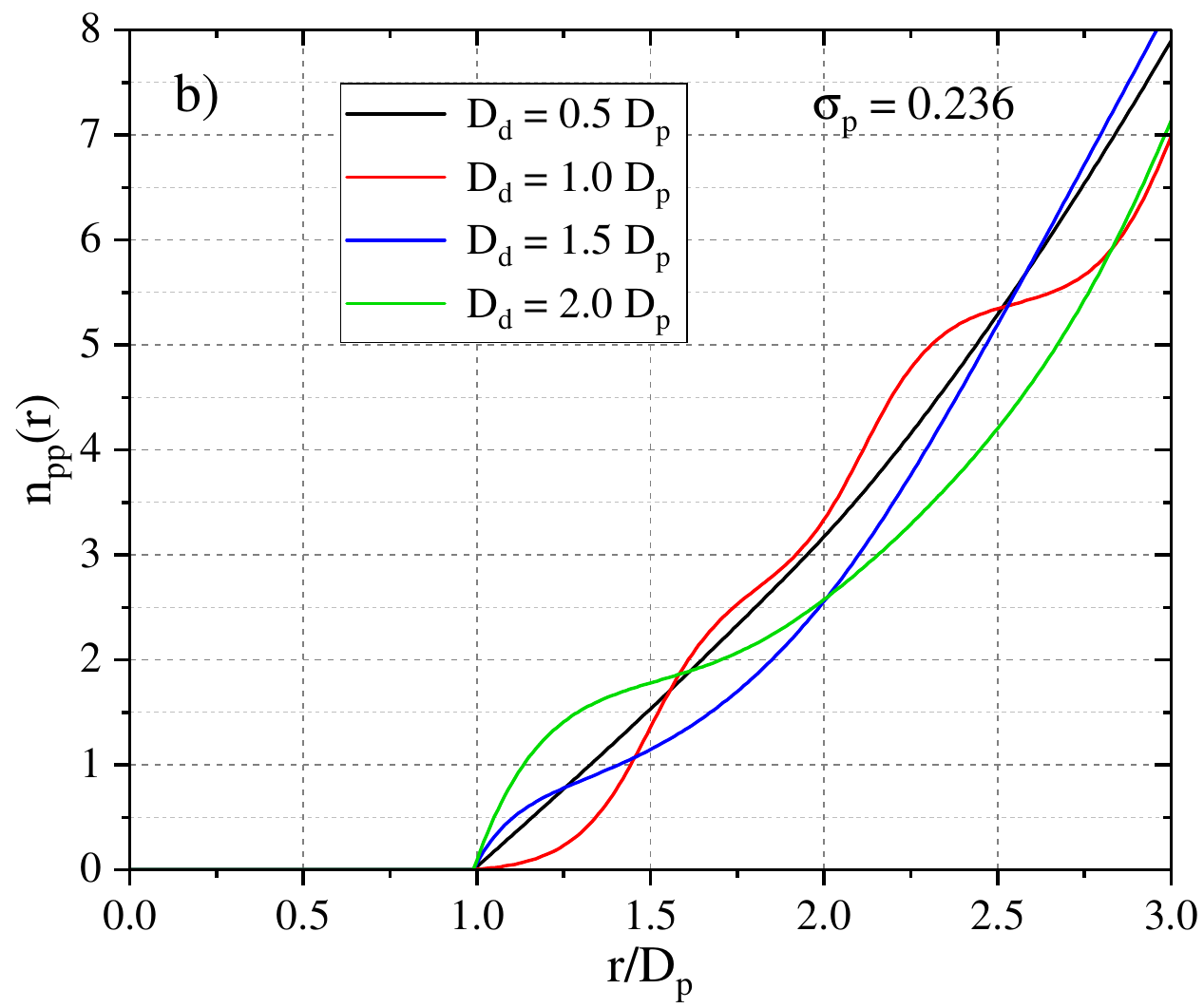}    
	\caption{Radial distribution function $g_{\rm pp}(r)$~(a) and cumulative coordination number $n_{\rm pp}(r)$~(b) for particle--particle pairs at different domain sizes $D_{\rm d}$ and fixed domain surface coverage $\sigma_{\rm d}=0.349$. The domains are arranged on a square lattice, and the particle surface coverage is $\sigma_{\rm p}=0.236$.}
	\label{fig:gpp}
\end{figure}

For completeness, Fig.~\ref{fig:gpp} presents the radial distribution functions $g_{\rm pp}(r)$ and the corresponding cumulative coordination numbers $n_{\rm pp}(r)$ for particle--particle pairs under the same conditions. These functions characterize the lateral ordering of the adsorbed particles and show how the domain pattern affects correlations between particles. As expected for hard disks, $g_{\rm pp}(r)$ is zero at distances smaller than the particle diameter. The first non-zero values appear near contact, $r/D_{\rm p} = 1$, where the behaviour depends strongly on the domain size. For $D_{\rm d}/D_{\rm p}=0.5$, the particle--particle correlations are relatively smooth. This is consistent with the fact that small domains create an effectively averaged attractive field, leading to a comparatively homogeneous particle distribution on the surface.
For $D_{\rm d}/D_{\rm p}=1.0$, the function $g_{\rm pp}(r)$ exhibits more pronounced oscillations. These oscillations reflect the ordered arrangement of the domains and the tendency of particles to occupy positions close to individual domain centers. In this case, the lateral structure of the particle layer is strongly influenced by the commensurability between the particle size and the domain size.
For larger domains, $D_{\rm d}/D_{\rm p}=1.5$ and $2.0$, the first peak of $g_{\rm pp}(r)$ near contact becomes more pronounced. This indicates that particles tend to be located closer to one another within the same enlarged adhesive region. In other words, larger domains promote local particle accumulation, because several particles can be adsorbed on or near the same domain. At the same time, since the number of domains decreases at fixed $\sigma_{\rm d}=0.349$, the resulting particle distribution becomes more heterogeneous: regions with enhanced particle concentration coexist with regions that are less populated.

The cumulative coordination number $n_{\rm pp}(r)$ in Fig.~\ref{fig:gpp}b supports this interpretation. At short distances above contact, the curves differ noticeably, showing that the local particle environment depends on the domain size. At larger distances, the curves become closer to each other, indicating that the differences are mainly local and are associated with the arrangement of particles within and around individual adhesive domains. Thus, the particle--particle correlations confirm that the domain size controls not only the adsorption capacity of individual domains but also the local structure of the adsorbed layer.

The results presented in Figs.~\ref{fig:gdp} and \ref{fig:gpp} show that the structural organization of adsorbed particles is determined by a balance between particle--domain attraction and steric constraints between particles. Small domains produce a relatively smooth and weakly localized adsorption pattern. Particle-sized domains, $D_{\rm d}/D_{\rm p}=1.0$, favour a one-particle--one-domain adsorption mechanism and lead to strong domain--particle correlations. Larger domains, $D_{\rm d}/D_{\rm p}=1.5$ and $2.0$, can accommodate several particles per domain, resulting in stronger local particle accumulation and more pronounced particle--particle correlations. These structural observations provide direct support for the interpretation of the adsorption isotherms and snapshots discussed above.

\section{Conclusions}

In this work, we studied the equilibrium adsorption of disk-like particles on patterned adhesive surfaces using Monte Carlo simulations. The surface was represented by fixed adhesive domains arranged either in an ordered or disordered pattern, while the particles interacted attractively with the domains and were subject to hard-core exclusion.

The results show that adsorption on patterned adhesive surfaces is controlled not only by the total attractive area but also by the geometry of the surface pattern. In particular, the domain size, domain surface coverage, and spatial arrangement of the domains jointly determine the adsorption isotherms and the lateral organization of the adsorbed particles. Small domains produce an effectively averaged attractive field, whereas domains comparable in size to the particles create favourable adsorption sites and lead to a more pronounced particle--domain matching effect.

From the point of view of enhancing adsorption, the most efficient domain size depends on the adsorption regime. At low and intermediate particle coverages, adsorption is strengthened when the domain size is comparable to the particle size, because particles can maximize their overlap with individual adhesive domains. At higher coverages, however, this advantage becomes weaker once the most favourable adsorption positions are occupied. In this regime, smaller and more numerous domains may become more effective, since they provide a more spatially distributed attractive pattern and allow more particles to maintain partial contact with adhesive regions. Increasing the domain surface coverage generally promotes adsorption by increasing the availability of favourable adhesive regions.

The comparison between ordered and disordered patterns indicates that the overall adsorption behaviour is governed mainly by domain size and domain surface coverage. Spatial disorder has a weaker but still noticeable effect, because it creates a broader distribution of local environments and smooths out adsorption features associated with particle--domain commensurability.

The analysis of particle configurations and correlation functions supports this interpretation. Small domains do not strongly localize particles around individual domain centers, particle-sized domains favour a one-particle--one-domain adsorption mechanism, and larger domains allow several particles to be associated with the same adhesive region. Thus, the local structure of the adsorbed layer reflects the balance between particle--domain attraction, surface-pattern geometry, and steric constraints between particles.

These findings suggest that adsorption capacity and lateral particle organization can be tuned by controlling the size, coverage, and spatial arrangement of adhesive domains. This may be useful for the design of functional patterned substrates, selective adsorption surfaces, biosensor platforms, for colloidal assembly templates, and surfaces for controlled immobilization of nano- and microparticles, proteins and cells. The results may also be relevant to affinity-based cell-sorting strategies, in which patterned adhesive regions are used to control the attachment of cells with different sizes, contact geometries, or binding energies~\cite{Badenhorst2025}.

Further extensions of the present model may consider binary mixtures of particles with different affinities to the adhesive domains, particles of different sizes, and polydispersity in domain sizes, which would allow one to address more realistic scenarios of selective adsorption and particle sorting on patterned surfaces.

\section{Acknowledgments}

The work is supported by the STCU Grant 7115. Computer time for the reported simulations was provided by the Interdisciplinary Center for Computer Simulations (Lviv), which supported by the NRFU Grant No.~2023.05/0019.

\end{document}